\documentclass[a4paper,fleqn]{cas-dc}

\usepackage[numbers]{natbib}
\usepackage{float}
\usepackage{caption} 
\usepackage{placeins}  
\usepackage{graphicx} 
\usepackage{balance}
\usepackage{threeparttable}
\usepackage{subcaption}

\def\tsc#1{\csdef{#1}{\textsc{\lowercase{#1}}\xspace}}
\tsc{AB}
\tsc{MMR}
\tsc{AS}
\tsc{EHB}

\begin{document}
\let\WriteBookmarks\relax
\def\floatpagepagefraction{1}
\def\textpagefraction{.001}
\shorttitle{Finite Element Analysis of Kidney MRI}
\shortauthors{A.  Bhattacharjee et~al.}

\title [mode = title]{Finite Element Elastographic Analysis of Kidney MRI for Computational Assessment of Chronic Kidney Disease}                      



\author[1]{A.  Bhattacharjee}[type=editor,
                        auid=000,bioid=1,
                        orcid=0009-0006-6496-7896]
												
\ead{2006029@eee.buet.ac.bd}


\author[2]{M. M. Rahman}[type=editor,
                        auid=000,bioid=1,
                         orcid=0000-0002-5000-761X]

\ead{mazderunb@gmail.com}


\author[3]{A. Srivastava}[type=editor,
                        auid=000,bioid=1,
                        orcid=0000-0002-8109-1420]

\ead{asrivast@uic.edu}


\author[4,5]{E. H. Bhuiyan}[type=editor,
                        auid=000,bioid=1,
                        orcid=0000-0002-3098-1095]
												
\cormark[1]
\ead{bhuiyan@uic.edu}

													


\affiliation[1]{organization={Department of Electrical and Electronic Engineering},
                addressline={Bangladesh University of Engineering and Technology}, 
                city={Dhaka},
                postcode={1000}, 
                state={Dhaka},
                country={Bangladesh}}
								
\affiliation[2]{organization={Faculty of Computer Science},
                addressline={University of New Brunswick}, 
                city={Fredericton},
                postcode={E3B 5A3}, 
                state={NB},
                country={Canada}}

\affiliation[3]{organization={Division of Nephrology, Department of Medicine},
                addressline={University of Illinois Chicago}, 
                city={Chicago},
                postcode={60612}, 
                state={IL},
                country={USA}}	

\affiliation[4]{organization={Center for Magnetic Resonance Research},
                addressline={Department of Radiology, University of Illinois Chicago}, 
                city={Chicago},
                postcode={60612}, 
                state={IL},
                country={USA}}	

\affiliation[5]{organization={AI.Health4All Center Initiatives, College of Medicine},
                addressline={University of Illinois Chicago}, 
                city={Chicago},
                postcode={60612}, 
                state={IL},
                country={USA}}	

\cortext[cor1]{Enamul Hoque Bhuiyan, Center for Magnetic Resonance Research, Department of Radiology and AI.Health4All Center Initiatives, University of Illinois Chicago, Chicago, IL 60612, USA, Phone: +1-203-214-6948, Fax: +1-312-355-1635, email: bhuiyan@uic.edu}


\begin{abstract}
\noindent
\textbf{Background and Objective:} Bridging kidney MRI with patient-specific computational biomechanics may enable accurate, noninvasive assessment of chronic kidney disease. To develop a finite element method (FEM)-based computational elastography framework for estimating renal biomechanical properties from conventional kidney MRI and to evaluate its ability to differentiate chronic kidney disease (CKD) from healthy controls.

\vspace{1em}

\noindent
\textbf{Methods:} Kidney segmentation masks from healthy control and CKD subjects were converted into triangulated surface meshes, and watertight surfaces were tetrahedralized to create volumetric FEM meshes. Kidney tissue was modeled as a nearly incompressible Kelvin–Voigt viscoelastic material. Harmonic wave propagation simulations were performed using FEM, followed by Helmholtz-based elastographic inversion to estimate complex shear modulus distributions. Derived biomechanical parameters were used for CKD classification using machine learning models.

\vspace{1em}

\noindent
\textbf{Results:} FEM-derived stiffness estimates showed consistency with stiffness ranges previously reported in human renal MRE studies and also showed measurable differences between healthy and CKD groups. Machine learning models trained on FEM-derived features achieved classification performance up to 86.26\% cross-validation accuracy with an AUC of 0.92.

\vspace{1em}

\noindent
\textbf{Conclusion:} FEM-derived elastographic parameters showed potential for distinguishing CKD from healthy tissue and may provide an alternative for renal biomechanical assessment without dedicated MRE acquisition.

\end{abstract}

\begin{graphicalabstract}
\includegraphics[width=1.0\textwidth]{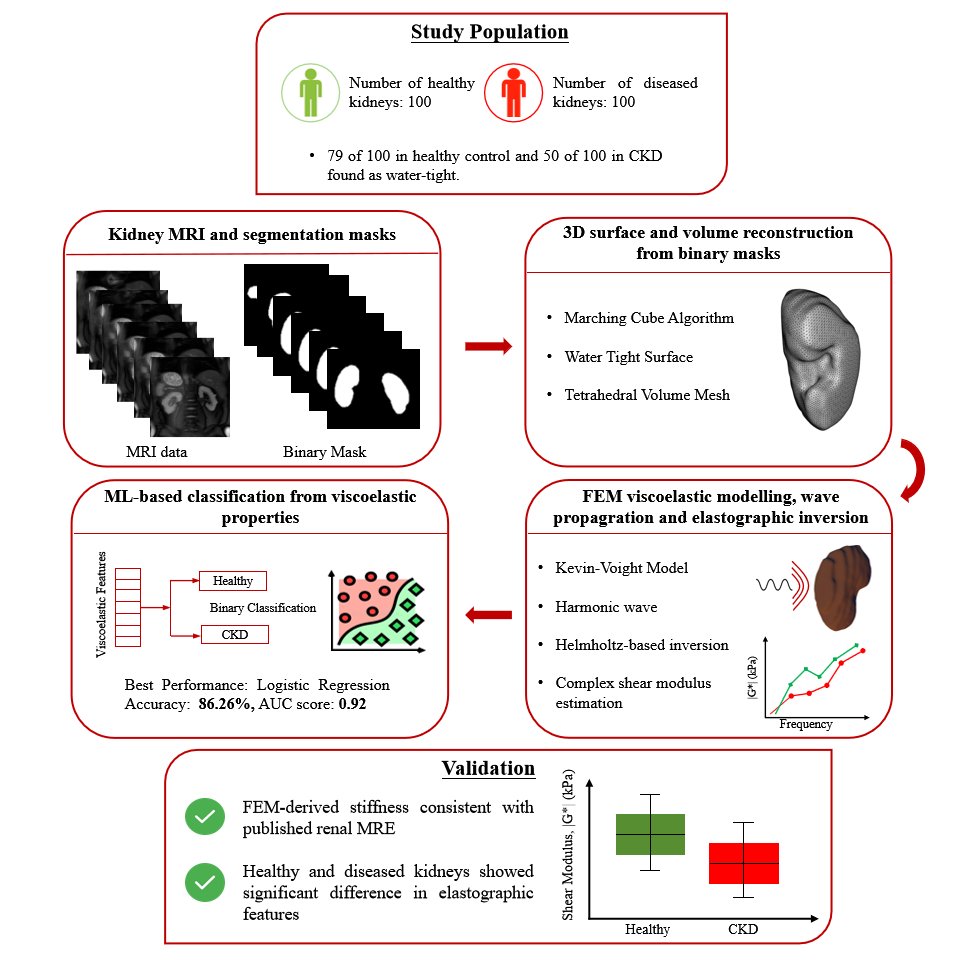}
\end{graphicalabstract}

\begin{highlights}
  \item Patient-specific kidney MRI was converted into watertight FEM meshes.
  \item Kelvin-Voigt FEM and Helmholtz inversion estimated renal viscoelasticity.
  \item FEM-derived stiffness differentiated healthy kidneys from CKD.
  \item Logistic regression achieved 0.92 AUC and 86\% cross-validated accuracy.
  \item The framework enables computational renal elastography without dedicated MRE.
\end{highlights}


\maketitle

\section{Introduction}

Chronic kidney disease (CKD) is a progressive disorder characterized by structural and functional deterioration of renal tissue. It remains a major global health burden, with about one in ten people having CKD~\cite{kearney2005global}. CKD is associated with several health complications like anaemia, bone disease, and increased risk of cardiovascular disease and even cancer~\cite{webster2017chronic}. Standard diagnosis of CKD includes renal biopsy, which has limitations due to its invasive nature~\cite{berchtold2017diagnosis}. Non-invasive methods include ultrasound and magnetic resonance imaging-based techniques. Ultrasound-based elastography techniques have been evaluated for diagnosing fibrosis in the liver and kidneys~\cite{gandhi2019magnetic}, but have some shortcomings. They only provide 1D stiffness of organs, have lower penetration depth in obese patients, and are limited by transducer force~\cite{gennisson2013ultrasound}. Many of the limitations associated with ultrasound-based elastography can be mitigated by magnetic resonance elastography (MRE). MRE measures externally generated harmonic shear waves using vibration-synchronized phase-contrast MRI. It reconstructs tissue mechanical properties through elastographic inversion algorithms~\cite{sack2023magnetic}. MRE provides volumetric assessment of tissue biomechanics with improved penetration depth, reduced operator dependence, and superior soft-tissue contrast~\cite{ajmera2022liver}. Progressive fibrosis, altered perfusion, and tissue remodeling associated with CKD can cause measurable changes in the mechanical properties of the kidney, including stiffness and viscoelastic behavior~\cite{inoue2011noninvasive}, which can be detected by MRE~\cite{brown2020utility}. However, the interpretation of renal stiffness remains complex because renal tissue exhibits nonlinear, anisotropic, and perfusion-dependent mechanical behavior~\cite{stabinska2024probing, gandhi2020magnetic}. Besides, respiratory motion, deep anatomical positioning, and the need for specialized acquisition hardware and pulse sequences also make the use of MRE difficult~\cite{mariappan2010magnetic}. 

Computational elastography offers a potential alternative by enabling physics-based estimation of tissue mechanical behavior without requiring dedicated hardware~\cite{bhuiyan2026mri}. Recent advances in computational biomechanics and finite element analysis (FEA) have enabled the modeling of anatomical structures from medical imaging data. FEA-based elastographic modeling has been widely investigated in soft tissue mechanics~\cite{sack2023magnetic, babaei2021magnetic}, wave propagation analysis~\cite{tomita2018numerical}, and inverse elasticity problems~\cite{mcgarry2022mapping, doyley2012model, alshareef2021integrating}. However, its application as a computational surrogate for renal elastography remains limited. 

\begin{figure*}
    \centering
    \includegraphics[width=0.8\linewidth]{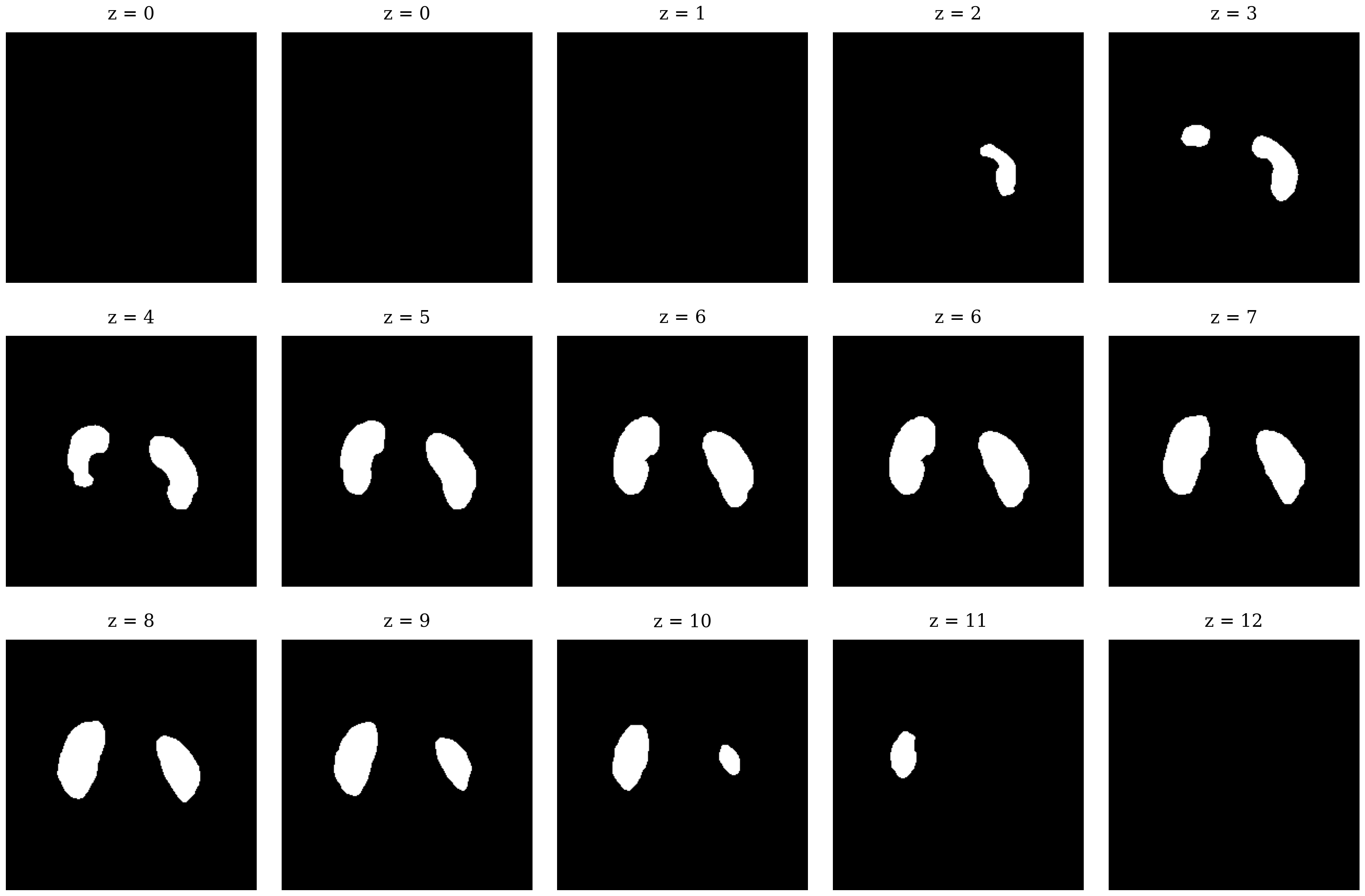}
    \caption{Representative 2D cross-sectional slice configurations of segmented human renal parenchyma from the NIfTI metadata}
    \label{fig:masks}
\end{figure*}

Patient-specific FEA pipelines commonly involve medical image segmentation, surface reconstruction using algorithms such as marching cubes~\cite{lorensen1998marching}, volumetric tetrahedral meshing, and constitutive material modeling. In elastographic inversion problems, Helmholtz-based formulations have been widely used to recover mechanical parameters from harmonic displacement fields~\cite{fushimi2022novel}. These methods have shown effectiveness in several organs, particularly in the brain~\cite{mittal20243d, wang2025data} and liver~\cite{lv2015study, yuan2010novel} studies. In parallel, machine learning and deep learning techniques have increasingly been explored for CKD detection using medical imaging data~\cite{anand2025optimized}. Convolutional neural networks and volumetric learning frameworks have demonstrated strong classification performance~\cite{islam2024classification, anoch2025detection}. Several recent works have also proposed advanced computational viscoelastic formulations~\cite{tomita2018numerical, qiang2015modeling}, including poroelastic and physics-informed neural-network frameworks~\cite{yin2025physics} for elastography. Although these studies included kidney tissue in broader multi-organ demonstrations~\cite{lim2023role}, they did not investigate finite-element reconstruction pipelines derived from conventional MRI nor evaluate disease classification using FEA-derived biomechanical descriptors.

We propose a computational renal elastography framework based on finite element modeling and viscoelastic inversion of conventional kidney MRI. The proposed pipeline reconstructs a patient-specific 3D kidney geometry and corresponding volumetric finite-element discretization and performs dynamic viscoelastic simulations using a Kelvin-Voigt~\cite{fung2013biomechanics, sack2023magnetic} material formulation. Biomechanical parameters were estimated using Helmholtz-based inversion and evaluated for CKD classification with machine learning.

We hypothesize that FEM-derived biomechanical descriptors from conventional kidney MRI capture disease-related alterations that distinguish healthy from CKD kidneys without dedicated MRE, enabling computational biomechanical assessment using routine renal MRI.

\section{Methodology}\label{sec2}

To create the 3D model of kidneys, the dataset from Daniel et al.~\cite{daniel2021automated} was used, which comprises abdominal magnetic resonance imaging (MRI) scans from a cohort of 60 human subjects, split equally between 30 healthy controls (HC) and 30 patients diagnosed with CKD spanning stages 1 to 5. Ground-truth segmented binary masks of the left and right kidneys were manually traced by trained expert observers. The segmented masks across different sequential depths are shown in Figure~\ref{fig:masks}. The structural and spatial constraints were extracted directly from the dataset's NIfTI headers, which reveal a spatial matrix dimension of $\text{dim} = [240 \times 240 \times 13]$ stored in double-precision floating-point ($\texttt{float64}$) format. The corresponding voxel properties are defined by an in-plane spatial resolution of $\Delta x \times \Delta y = 1.458\text{ mm} \times 1.458\text{ mm}$ and a slice thickness of $\Delta z = 5.5\text{ mm}$, yielding a physical voxel volume of $\sim 11.69\text{ mm}^3$. Matrix-to-physical space orientation is preserved using the s-form coordinate transformations, parameterized by the directional mapping vectors $\mathbf{s}_x = [1.458, 0, 0, 0]^T$, $\mathbf{s}_y = [0, 0, -5.5, 0]^T$, and $\mathbf{s}_z = [0, 1.458, 0, 0]^T$. Because this dataset intentionally spans a highly diverse morphological range, the derived 3D anatomical masks provide a robust, physically grounded foundation for reconstruction of a 3D kidney model and elastographic computations.

The first stage of the computational pipeline involved reconstructing a geometrically consistent surface mesh from the segmented masks of kidney MR images. Since medical imaging data is inherently anisotropic, with voxel spacing differing along spatial axes, the image was resampled and binarized to a uniform resolution. 

\subsection{Surface Extraction from kidney masks}

A connected component labeling operation was applied to identify distinct anatomical regions. 
Only the largest components satisfying the condition 
were retained. 
This step isolates the kidneys while suppressing noise and spurious regions. To mitigate voxel-level discontinuities and staircase artifacts, Gaussian smoothing was applied to the binary volume.
The smoothed scalar field was converted into a triangulated surface using the marching cubes algorithm~\cite{lorensen1998marching}. 
Vertices were transformed from voxel coordinates to world coordinates using the affine transformation, $\mathbf{v}_i^{(w)} = \mathbf{A} \cdot \mathbf{v}_i^{(v)}$, where $\mathbf{A}$ is the NIfTI affine matrix.

\begin{figure*}[t]
    \centering
    \includegraphics[width=0.44\linewidth]{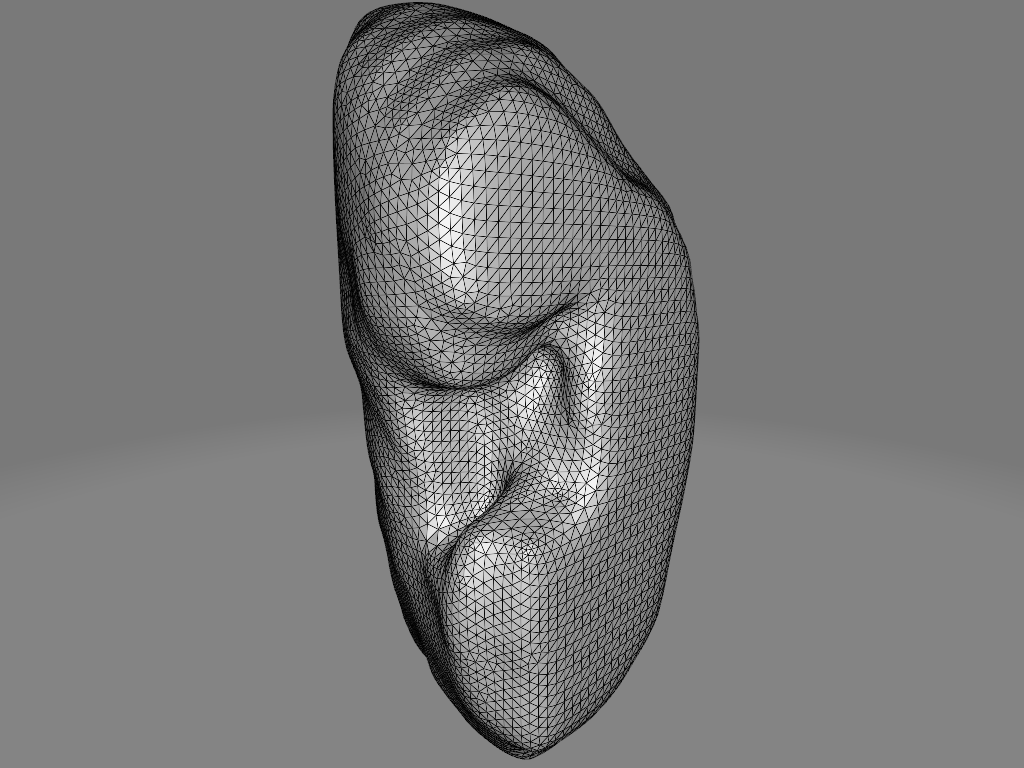}
    \quad
    \includegraphics[width=0.44\linewidth]{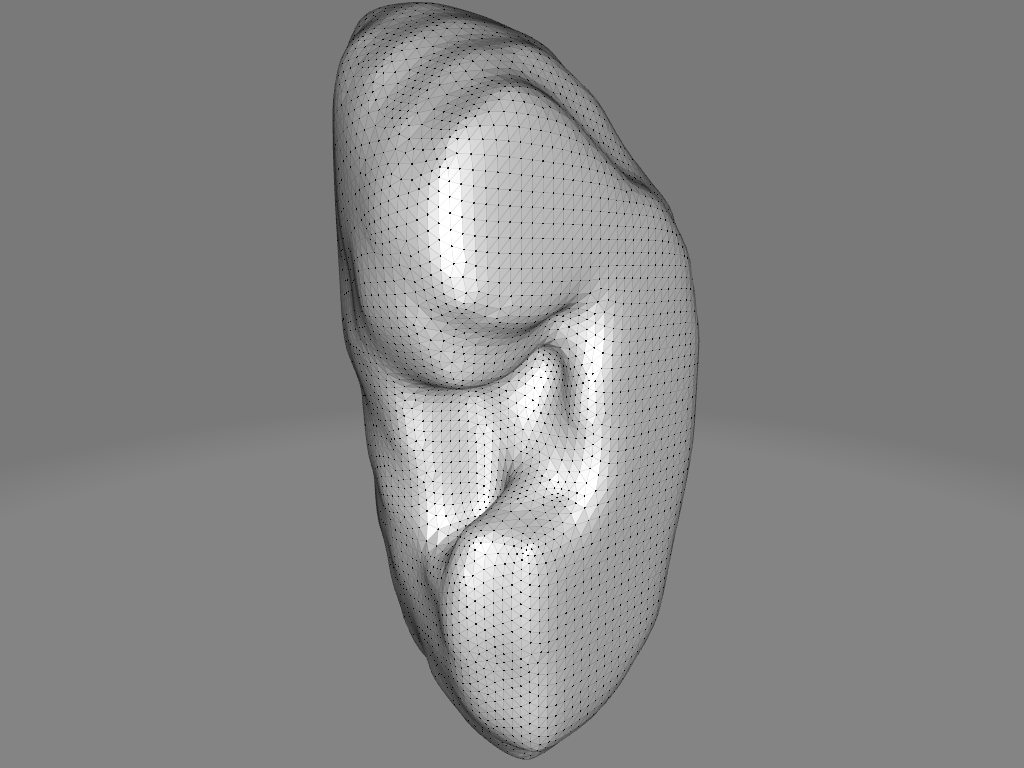}
    \caption{Reconstructed 3D surface mesh topology generated from the segmented binary clinical masks showing triangular faces and the vertices of the kidneys.}
    \label{fig:surface_mesh}
\end{figure*}

The reconstructed surface mesh was cleaned by removing non-manifold elements, duplicate vertices, and degenerate faces. Vertices not associated with any face were eliminated. Topological integrity was calculated using edge-incidence counts.  To further improve geometric quality without inducing volumetric shrinkage, Taubin smoothing~\cite{taubin1995signal} was applied to mitigate the effect of discontinuities in the surface introduced by corrective operations. 



Finally, the enclosed volume of the triangulated surface was computed using the divergence theorem:

\begin{equation}
V = \frac{1}{6} \sum_{f \in \mathcal{F}} (\mathbf{v}_i \times \mathbf{v}_j) \cdot \mathbf{v}_k
\end{equation}

where $(i,j,k)$ are vertices of each triangular face. The result is converted to milliliters for anatomical interpretation. The reconstructed triangular surfaces and the vertices of a kidney is shown in Figure~\ref{fig:surface_mesh}. Although the dataset comprised 30 healthy and 30 CKD patients (60 kidneys), patients 26–30 had multiple segmentation masks. So, all masks were simulated, resulting in a total of 100 kidneys in both healthy and diseased cases.

\subsection{Volumetric Mesh Generation}
After the construction of a topologically consistent surface mesh, volumetric discretization was conducted for finite element analysis (FEA). This process converted the closed triangulated surface into a tetrahedral mesh, which served as the computational domain $\Omega \subset \mathbf{{R}}^3$ over which governing equations were solved. Before volumetric meshing, the integrity of the surface mesh was rigorously validated, and several topological invariants were computed, such as the Euler characteristic ($\chi = 2$), edge incidence, and water-tightness.


Only meshes satisfying these constraints proceeded to volumetric meshing. If the surface failed validation, a conservative repair pipeline was applied, such as small boundary loops were closed to enforce watertightness and faces with near-zero area were eliminated. Identical triangles were removed using canonical sorting, and vertices within a tolerance 
were merged.
The repaired surface mesh was imported into the meshing framework, where geometric classification was conducted. The surface was partitioned into patches based on a feature angle, $\theta$, based segmentation. 
Edges with angles exceeding $\theta$ were treated as feature boundaries, and multiple feature angles were tested iteratively $\theta \in \{5^\circ, 10^\circ, 20^\circ, 30^\circ, 40^\circ \}$. This adaptive strategy reduced failures in geometric parameterization, particularly for complex anatomical surfaces.

The volumetric mesh was generated using a Delaunay-based tetrahedralization algorithm ~\cite{delaunay1934sphere, shewchuk1998tetrahedral}. The domain $\Omega$ was discretized into a set of tetrahedral elements, 
where each tetrahedron 
was defined by four vertices,
Mesh resolution was controlled by size constraints, $h_{\min} \leq h(\mathbf{x}) \leq h_{\max}$, where $h(\mathbf{x})$ denotes the characteristic edge length of a specific element $T_k$. In this implementation, $h_{\min} = 0.5 \text{mm},~ h_{\max} = 3.0 \ \text{mm}$ was used. 
First-order elements 
were used for computational efficiency. Element quality was evaluated using the Signed Inverse Condition Number (SICN), which measured distortion $\text{SICN}(T_k) \in [-1,1]$. Values close to 1 indicate well-shaped elements, while values near 0 indicate degeneracy. A warning threshold was defined ($\text{SICN}_{\min} < 0.1$) to denote poor-quality elements. Besides, tetrahedral volume was compared with surface-derived volume to check consistency. These ensure that the discretization was suitable for stable FEA computations. Note that the left and right kidneys were separated for volumetric mesh creation. 
\begin{table*}[htbp]
\centering
\caption{Material Properties and Simulation Constants Implemented in the Finite Element Analysis}
\label{tab:fem_constants}
\begin{tabular}{p{0.35\textwidth}p{0.15\textwidth}p{0.25\textwidth}}
\toprule
\textbf{Physical Parameter Description} & \textbf{Symbol} & \textbf{Value} \\ \midrule
\multicolumn{3}{l}{\textit{\textbf{Solid Mechanics and Boundary Stiffness Properties}}} \\
Volumetric Mass Density & $\rho$ & $1050.0$ $\text{kg/m}^3$ 
\\
Young's Modulus of Elasticity & $E$ & $9000.0$ $\text{Pa}$ 
\\
Poisson's Ratio & $\nu$ & $0.49$ (dimensionless) 
\\
Dynamic Viscosity (Kelvin-Voigt) & $\eta$ & $4.0$ $\text{Pa}\cdot\text{s}$ \\
Winkler Foundation Bed Stiffness & $k_w$ & $15000.0$ $\text{N/m}^3$ \\ \midrule 


\multicolumn{3}{l}{\textit{\textbf{Harmonic MRE Dynamic \& Time-Stepping Parameters}}} \\
Dynamic MRE Excitation Frequency & $f$ & $50.0$ $\text{Hz}$ \\
Boundary Displacement Amplitude & $U_0$ & $1.0\times 10^{-4}$ $\text{m}$ ($0.1$ mm) \\
Incremental Time Step ($\Delta t$) & $\Delta t$ & $1.0\times 10^{-3}$ $\text{s}$ ($1.0$ ms) \\
\bottomrule
\end{tabular}
\end{table*}

\subsection{Finite Element Analysis and Elastographic Inversion}

The kidney was modeled as a nearly incompressible, viscoelastic solid. The constitutive behavior followed a Kelvin-Voigt model~\cite{fung2013biomechanics}, where stress was decomposed into elastic and viscous components:
\begin{equation}
\sigma = \sigma_{\text{elastic}} + \sigma_{\text{viscous}}
\end{equation}

The elastic stress tensor was modeled as linear isotropic elasticity:
\begin{equation}
\sigma_{\text{elastic}} = \lambda (\nabla \cdot \mathbf{u}) \mathbf{I} + 2\mu \boldsymbol{\varepsilon}(\mathbf{u})
\end{equation}

where $\mathbf{u}$ is the displacement vector ($\mathbf{u} = [u_x, u_y, u_z]^T$), $\mathbf{I}$ denotes the second-order identity tensor used to scale the volumetric dilation ($\nabla \cdot \mathbf{u}$) across all spatial dimensions equally and $\lambda$, $\mu$ are Lamé parameters. 


The strain tensor was defined as $\varepsilon(\mathbf{u}) = \frac{1}{2}\left(\nabla \mathbf{u} + (\nabla \mathbf{u})^T \right)$. Finally, the viscous contribution was modeled as:
\begin{equation}
\sigma_{\text{viscous}} = 2\eta \varepsilon(\dot{\mathbf{u}})
\end{equation}
 where $\eta$ is the dynamic viscosity and $\varepsilon(\dot{\mathbf{u}})$ is the strain rate tensor. 

\paragraph{Harmonic Viscoelastic Wave Propagation:}

The dynamic problem models shear wave propagation induced by harmonic excitation (Magnetic Resonance Elastography regime). The governing equation:

\begin{equation}
\rho \ddot{\mathbf{u}} = \nabla \cdot \boldsymbol{\sigma}
\end{equation}

Substituting the Kelvin--Voigt model:
\begin{equation}
\rho \ddot{\mathbf{u}} =
\nabla \cdot \left[
\lambda (\nabla \cdot \mathbf{u}) \mathbf{I} +  2\mu \boldsymbol{\varepsilon}(\mathbf{u})
+ 2\eta \boldsymbol{\varepsilon}(\dot{\mathbf{u}})
  \right]
  \end{equation}

The time discretization follows the Newmark-$\beta$ mechanism~\cite{newmark1959method} 
with $\beta = 0.25$, $\gamma = 0.5$, ensuring unconditional stability and no introduction of artificial numerical damping. This ensured that any damping or phase lag observed in the steady-state snapshots resulted from physical material viscosity, not a numerical artifact of the solver.

\paragraph{Boundary Conditions:}

A harmonic displacement was applied at the superior surface $u_z(t) = A \sin(\omega t)$. Winkler-type absorbing boundary conditions were applied on lateral and inferior surfaces:
\begin{equation}
\boldsymbol{\sigma} \cdot \mathbf{n} + k_w \mathbf{u} = 0
\end{equation}

After reaching steady-state oscillation, the displacement field was transformed using the Fourier transform, which yields a complex displacement field, $\hat{u} = u_{\text{re}} + i u_{\text{im}}$. The viscoelastic properties were recovered using the Helmholtz equation~\cite{muthupillai1995magnetic}:
\begin{equation}
\nabla^2 \hat{u} + k^2 \hat{u} = 0
\end{equation}

where the complex wavenumber, $k$, is related to the complex dynamic shear modulus, $G^*$ via:
\begin{equation}
k^2 = \frac{\rho \omega^2}{G^*}, \quad G^* = G' + iG''
\end{equation}
Here, the parameter $G'$ denotes the storage modulus, representing the material's stiffness or rigidity, and $G''$ is the loss modulus, denoting the material's damping characteristics. Rearranging, we estimated the complex shear modulus, whose magnitude is the primary desired parameter in magnetic resonance elastography.
\begin{equation}
G^* = -\frac{\rho \omega^2 \hat{u}}{\nabla^2 \hat{u}}
\end{equation}

The Laplacian was computed using a weak formulation:
\begin{equation}
\int_{\Omega} (\nabla^2 u) v  d\Omega =  \int_{\Omega} \nabla u \cdot \nabla v  d\Omega - \int_{\partial \Omega} \frac{\partial u}{\partial n} v  d\Gamma
  \end{equation}

To suppress numerical noise, a Helmholtz-type smoothing filter was applied:
\begin{equation}
u_s - \sigma^2 \nabla^2 u_s = u
\end{equation}

This corresponds to Tikhonov regularization~\cite{fuhry2012new} in the $H^1$ space. For robustness, $G^*$ was estimated locally using neighborhood-based least squares:
\begin{equation}
\min_{G',G''} \left| \nabla^2 \hat{u} + \frac{\rho \omega^2}{G' + iG''} \hat{u} \right|^2
\end{equation}

The domain interior was restricted to avoid boundary artifacts, $z \in (z_{\min} + \lambda/4, z_{\max} - \lambda/4)$. Finally, a resolution constraint was applied where element size followed the condition, $\frac{\lambda}{h}\gtrsim5$, to accurately resolve shear waves. The magnitude and phase of the complex modulus and other relevant result parameters derived from this elastographic simulation were stored for binary classification between chronic and healthy kidneys.

\subsection{ML-based Classification}

The FEM-derived dataset was used as a binary classification problem to distinguish healthy kidneys from CKD-affected samples. The feature space consisted of mechanical descriptors, including shear moduli, modulus ratios, viscoelastic parameters, geometric scaling factors, and numerical resolution indicators. A range of supervised learning models was considered, spanning linear, kernel-based, ensemble, and neural approaches. Model performance was evaluated using repeated stratified cross-validation with five folds and ten repetitions. This approach ensured stable and statistically meaningful estimates in the presence of limited sample sizes. To address the class imbalance, Synthetic Minority Oversampling Technique (SMOTE)~\cite{chawla2002smote} was applied during model training. Note that SMOTE was applied exclusively within the training partition of each cross-validation fold using an imbalanced-learning-aware pipeline. It ensured that no synthetic samples derived from validation-fold neighbours were present during evaluation. Multiple evaluation metrics were computed, including accuracy, sensitivity, specificity, F1-score, and ROC-AUC. 

Each model was implemented within a pipeline that included feature standardization, ensuring consistent scaling across algorithms sensitive to feature magnitude. To explicitly quantify overfitting, the difference between training accuracy and cross-validation accuracy was computed for each model. This gap was used as a direct indicator of generalization performance. Hyperparameter optimisation was performed using Bayesian search via the Optuna framework~\cite{akiba2019optuna}, employing the Tree-structured Parzen Estimator (TPE) algorithm~\cite{bergstra2011algorithms} with 100 trials per model. The optimisation objective was the mean ROC-AUC computed over an inner five-fold stratified cross-validation, entirely nested within the outer repeated cross-validation loop to prevent any leakage of evaluation data into the tuning process. This nested structure ensured that reported generalisation metrics reflect performance on samples that were unseen not only during model fitting but also during hyperparameter selection.

\section{Results}\label{sec3}
\begin{figure}
    \centering
    \includegraphics[width=1\linewidth]{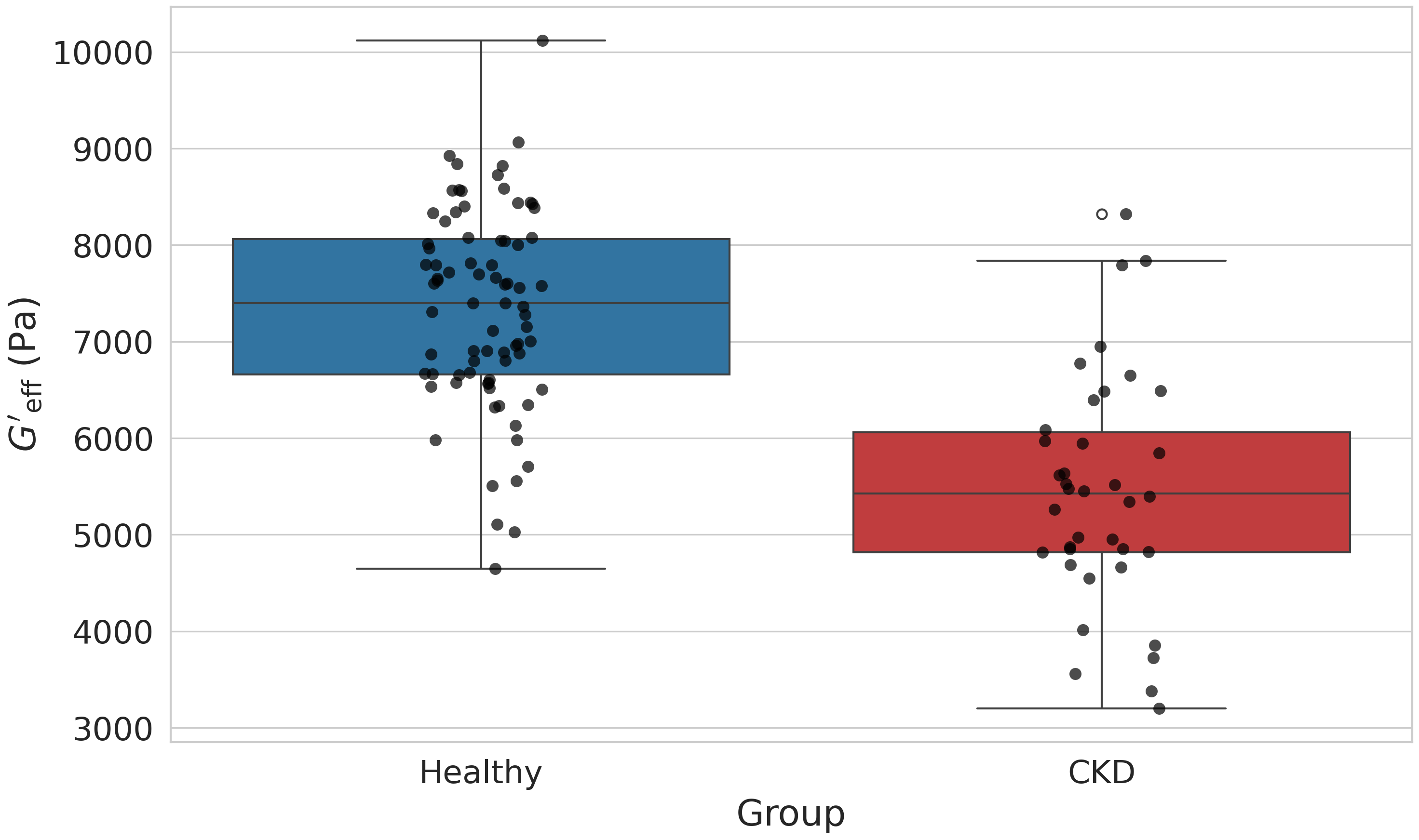}
    \caption{Box plot of effective storage modulus ($G'_{\text{eff}}$) demonstrating the distinct viscoelastic stiffness distribution between healthy control (HC) and chronic kidney disease (CKD) cohorts.}
    \label{fig:G'}
\end{figure}

As the primary target is to distinguish between healthy and diseased kidneys, we analyzed the statistical properties of FEA-derived parameters. Figure~\ref{fig:G'} shows the boxplot of the effective storage modulus for the two kidney groups, where a clear difference between the two groups can be seen. As CKD progresses, severe capillary rarefaction, ischemia, and loss of functional nephrons cause a dramatic drop in renal perfusion pressure, causing the organ to appear softer on elastography.

\begin{figure*}
    \centering
    \includegraphics[width=0.8\linewidth]{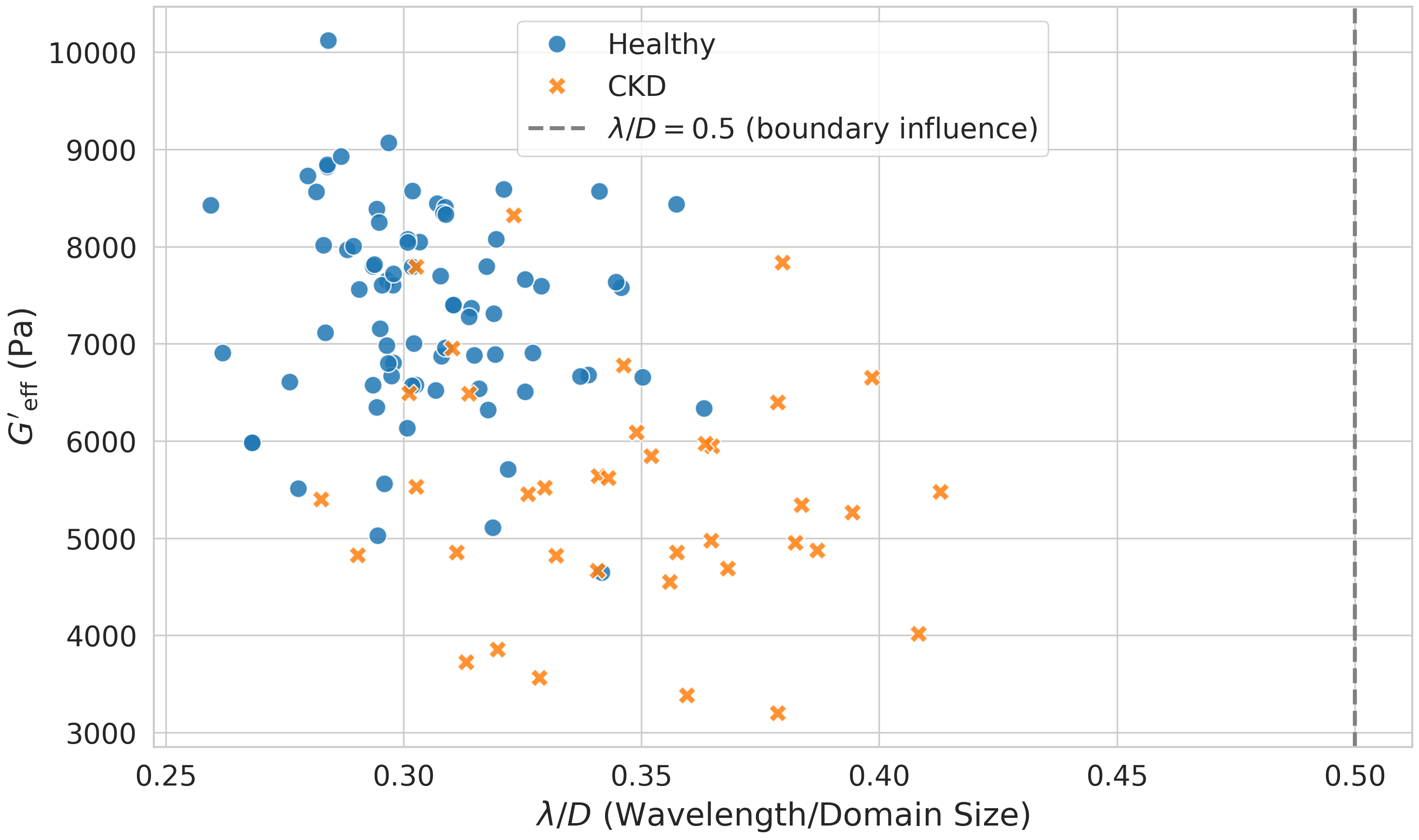}
    \caption{Scatter plot of effective storage modulus ($G'_{\text{eff}}$) as a function of the wavelength-to-domain ratio ($\lambda/D$), illustrating regional clustering and numerical stability bounds.}
    \label{fig:G'vslambda}
\end{figure*}

For a characteristic domain dimension, $D$, the ratio of $\lambda/D$ indicates the numerical stability of the inversion algorithm and the accuracy of derived parameters. A range of 0.2 to 0.5 provides enough spatial curvature for the finite element stencil to compute steady, non-zero derivatives ($\nabla^2 \hat{u}$) with high signal-to-noise ratios. Also, the waves preserve enough energy to travel from the boundary excitation point through to the deep medullary and cortical zones before being completely extinguished by material viscosity ($\eta$). Figure~\ref{fig:G'vslambda} shows the scatter plot of storage modulus with the $\lambda/D$ ratio, where healthy kidneys center around a $\lambda/D$ ratio of 0.3 and diseased kidneys center around 0.35. The obtained values aligned well with the desired range and also showed a distinct difference between healthy and diseased kidneys.

\begin{figure}
    \centering
    \includegraphics[width=0.9\linewidth]{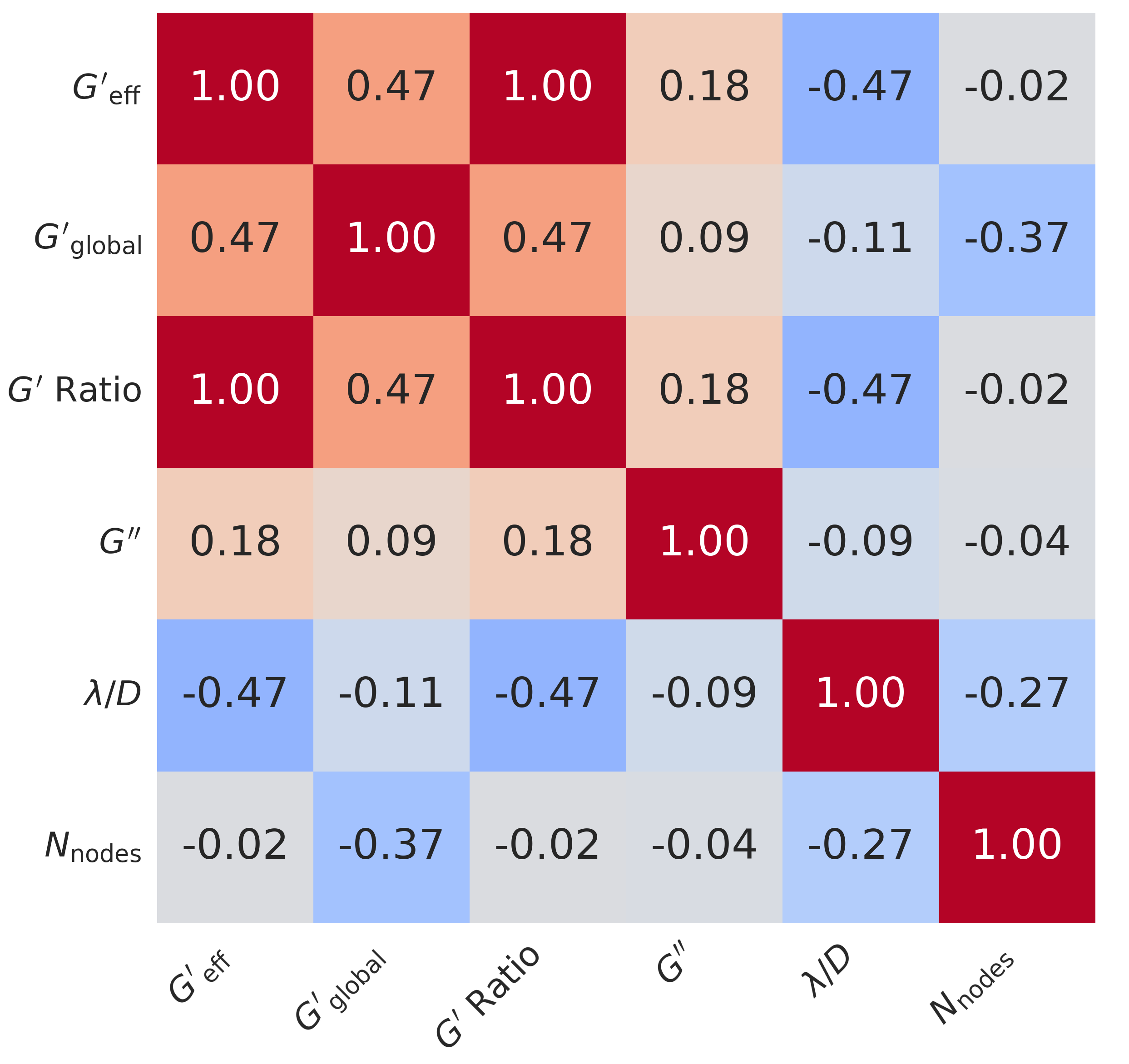}
    \caption{Pearson correlation coefficient ($r$) heatmap evaluating the linear dependencies among FEA-derived biomechanical descriptors and mesh parameters. Color intensity indicates the strength of the correlation, ranging from dark red (perfect positive correlation, $r = 1.00$) to dark blue (negative correlation).}
    \label{fig:corr_heatmap}
\end{figure}

\begin{figure}
    \centering
    \includegraphics[width=1\linewidth]{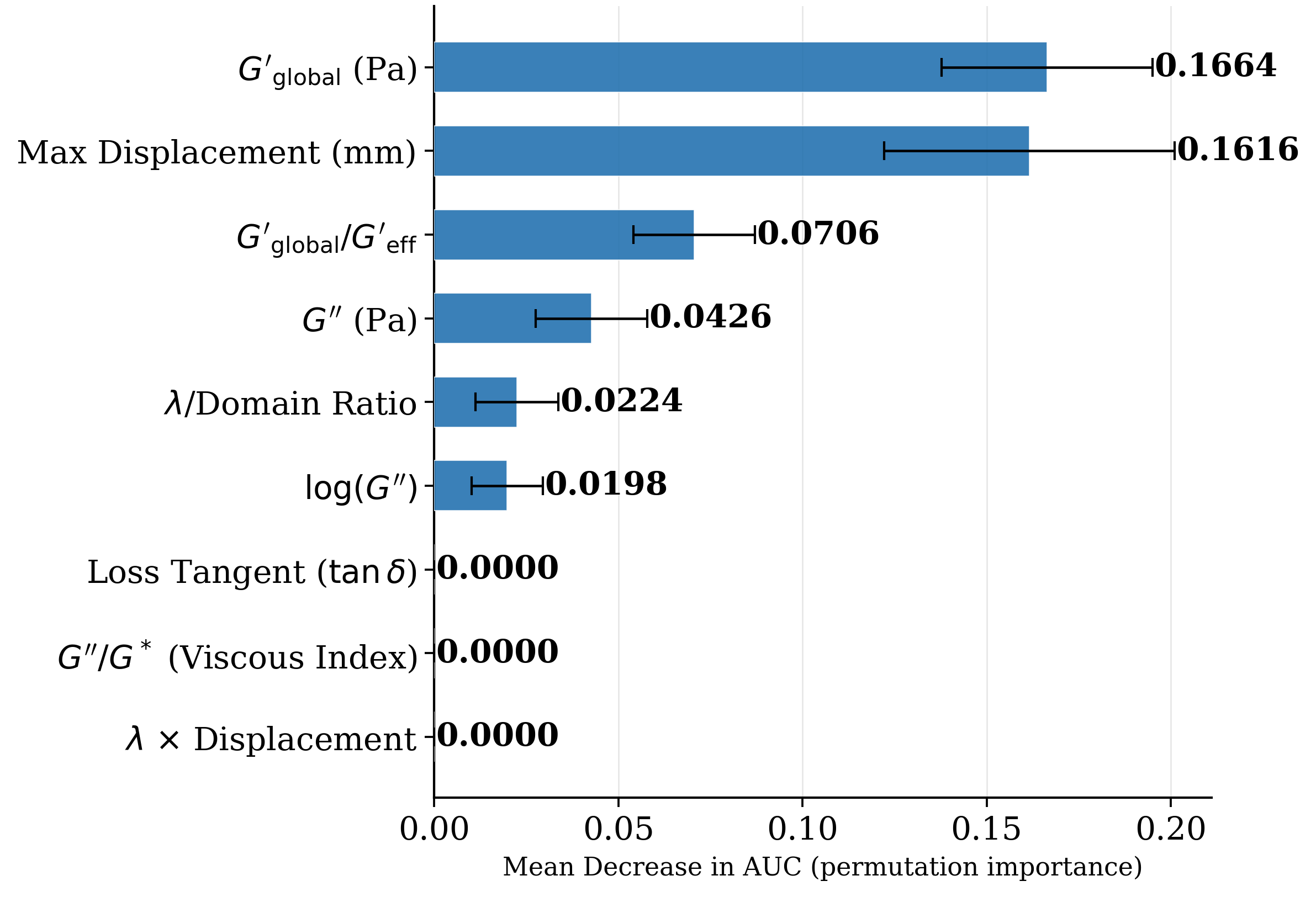}
    \caption{Feature relevance mapping via permutation importance, illustrating the performance degradation of the logistic regression model upon individual parameter shuffling}
    \label{fig:AUC permutation importance}
\end{figure}

A Pearson correlation analysis was performed to evaluate the linear relationships among the extracted biomechanical and numerical features. As shown in Figure~\ref{fig:corr_heatmap}, the effective storage modulus ($G'_{\text{eff}}$) and the $G'$ ratio exhibited a perfect positive correlation ($r = 1.00$), indicating complete statistical redundancy. Both $G'_{\text{eff}}$ and the $G'$ ratio demonstrated a moderate negative correlation with the wavelength-to-domain ratio ($\lambda/D$, $r = -0.47$), which is consistent with the physical coupling between increased tissue stiffness and longer acoustic wavelengths. In contrast, the loss modulus ($G''$) showed a weak correlation with $G'_{\text{eff}}$ ($r = 0.18$), confirming that the inversion pipeline successfully decouples the elastic and viscous components of the tissue. Finally, the total number of mesh nodes ($N_{\text{nodes}}$) showed a negligible correlation with both $G'_{\text{eff}}$ ($r = -0.02$) and $G''$ ($r = -0.04$), demonstrating that the recovered viscoelastic properties are independent of variations in spatial discretization density.

\begin{figure*}[!t]
\centering

\subfloat[Viscoelastic modulus versus excitation frequency.]
{
    \includegraphics[width=0.46\textwidth]
    {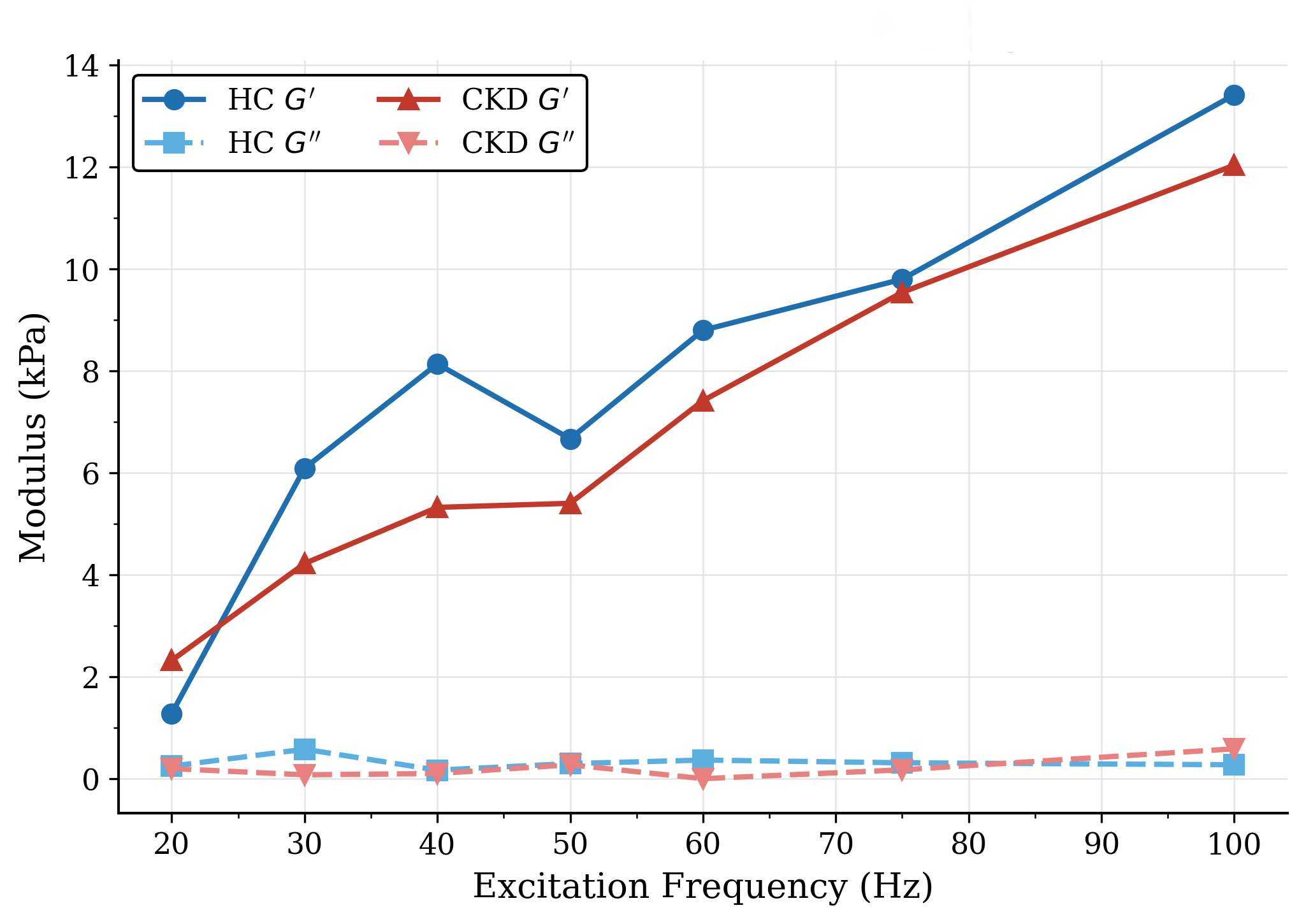}
    \label{fig:freq_sweep}
}
\hfill
\subfloat[Viscoelastic modulus versus Young's modulus, $E$.]
{
    \includegraphics[width=0.46\textwidth]
    {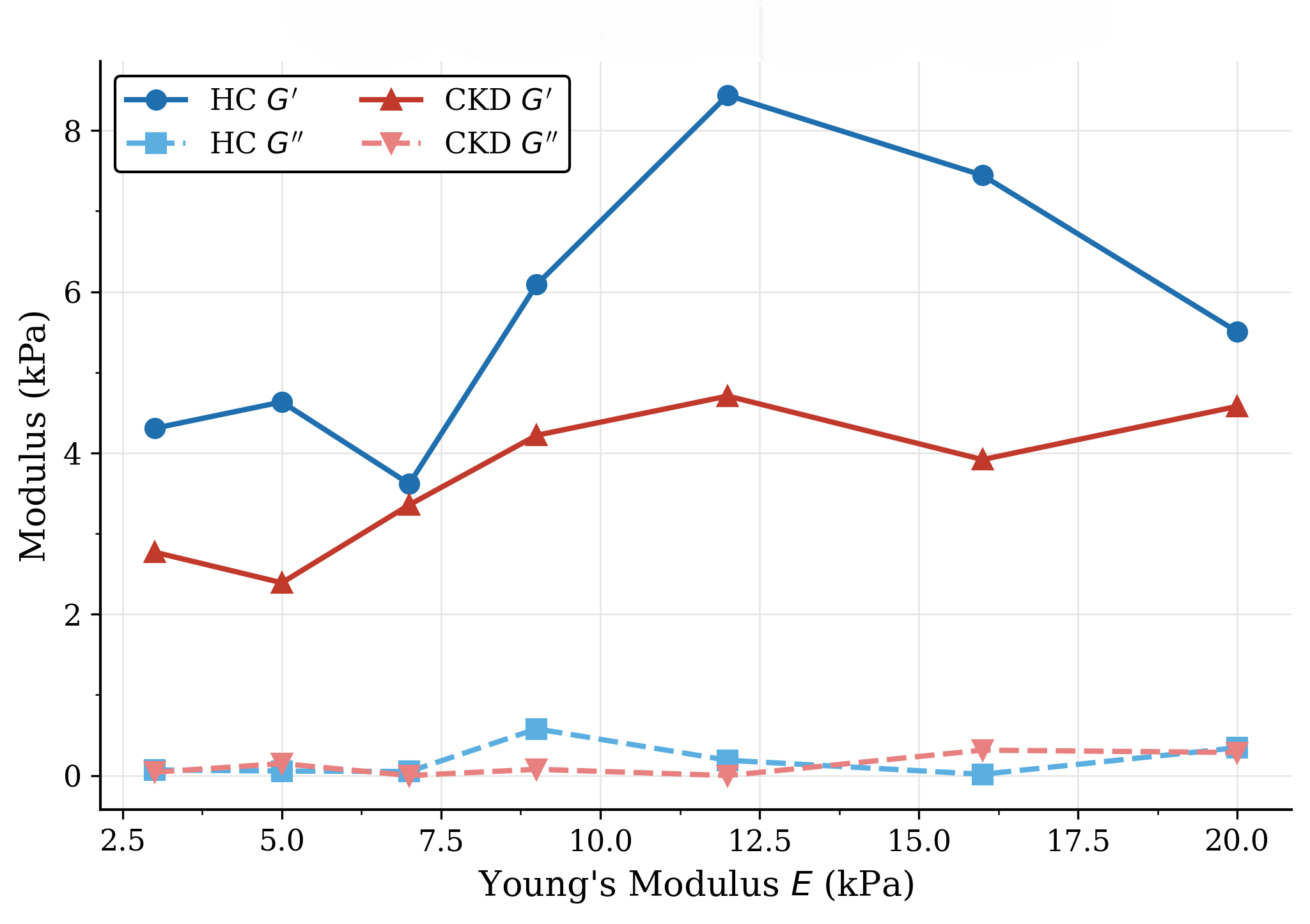}
    \label{fig:young_sweep}
}

\vspace{0.3cm}

\subfloat[Viscoelastic modulus versus Poisson's ratio, $\nu$.]
{
    \includegraphics[width=0.46\textwidth]
    {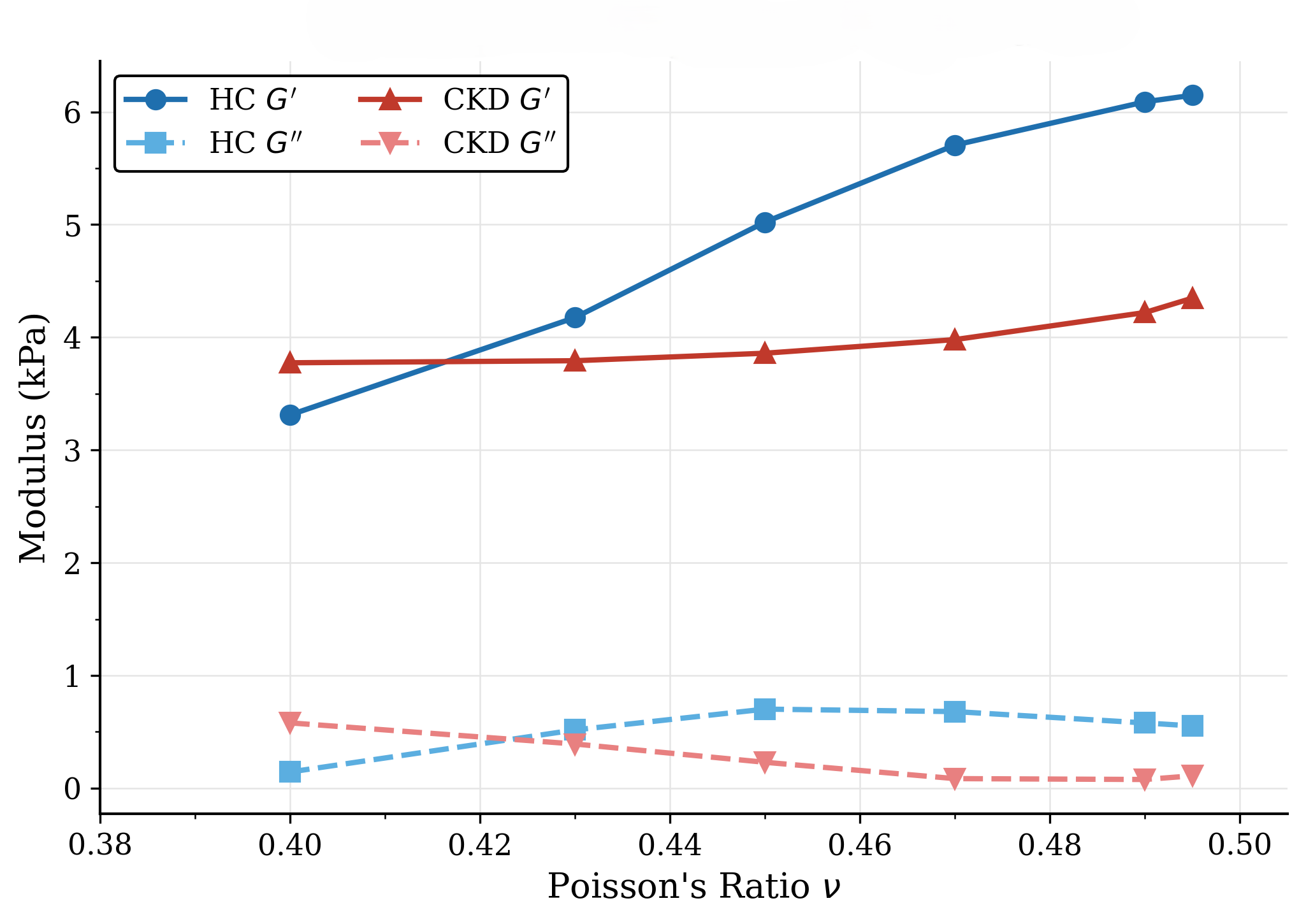}
    \label{fig:poisson_sweep}
}
\hfill
\subfloat[Viscoelastic modulus versus tissue density, $\rho$.]
{
    \includegraphics[width=0.46\textwidth]
    {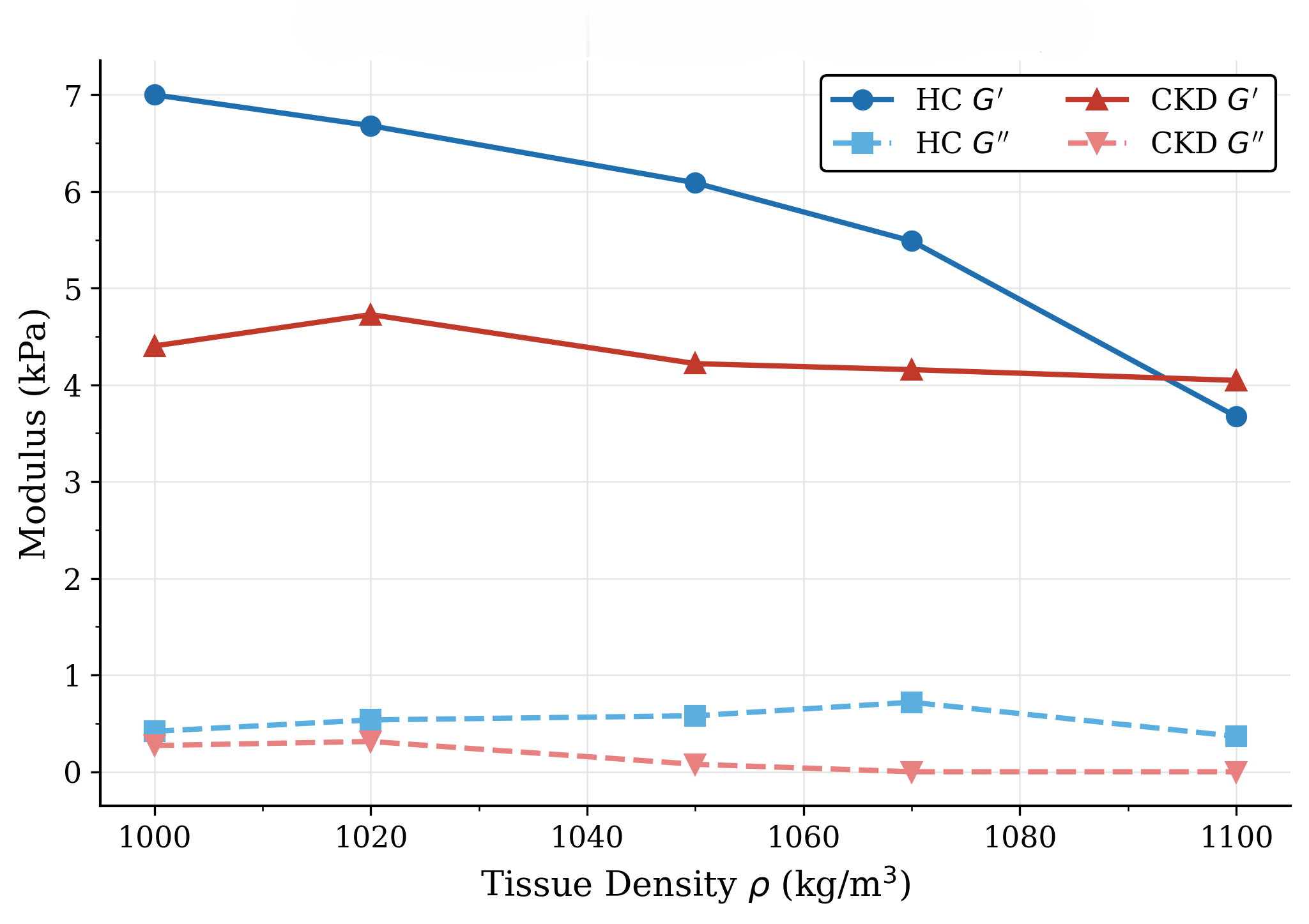}
    \label{fig:density_sweep}
}

\caption{Parameter-sweep analysis of viscoelastic moduli for healthy control and chronic kidney disease tissues.}
\label{fig:parameter_sweep}
\end{figure*}

Baseline values presented in Table~\ref{tab:fem_constants} were selected to fall within physiologically plausible ranges reported for renal tissue and soft-tissue MRE \cite{heymsfield1979accurate, liu2016classifying}. Their influence on recovered moduli was further assessed via the parameter sweep in Figure~\ref{fig:parameter_sweep}.  The parameter sweep analysis shows that the recovered viscoelastic moduli are most strongly influenced by excitation frequency and tissue stiffness, consistent with the frequency-dependent behavior of soft biological tissues reported in the literature~\cite{rouviere2011magnetic, low2015reliability}.  Sensitivity analysis demonstrated that the recovered viscoelastic parameters were primarily influenced by excitation frequency and tissue stiffness, while showing comparatively lower sensitivity to tissue density and Poisson's ratio. The weak sensitivity to tissue density supports the common assumption of constant tissue density in MRE, whereas the moderate sensitivity to Poisson's ratio near the incompressible limit highlights the importance of accurate compressibility characterization. These findings are consistent with previous MRE studies and support the numerical robustness of the proposed inversion framework for computational assessment of renal biomechanics. The parameter sweep analysis shows that the recovered viscoelastic moduli are most strongly influenced by excitation frequency and tissue stiffness, consistent with the frequency-dependent behavior of soft biological tissues reported in the literature~\cite{rouviere2011magnetic, low2015reliability}.

\begin{figure}
    \centering
    \includegraphics[width=1\linewidth]{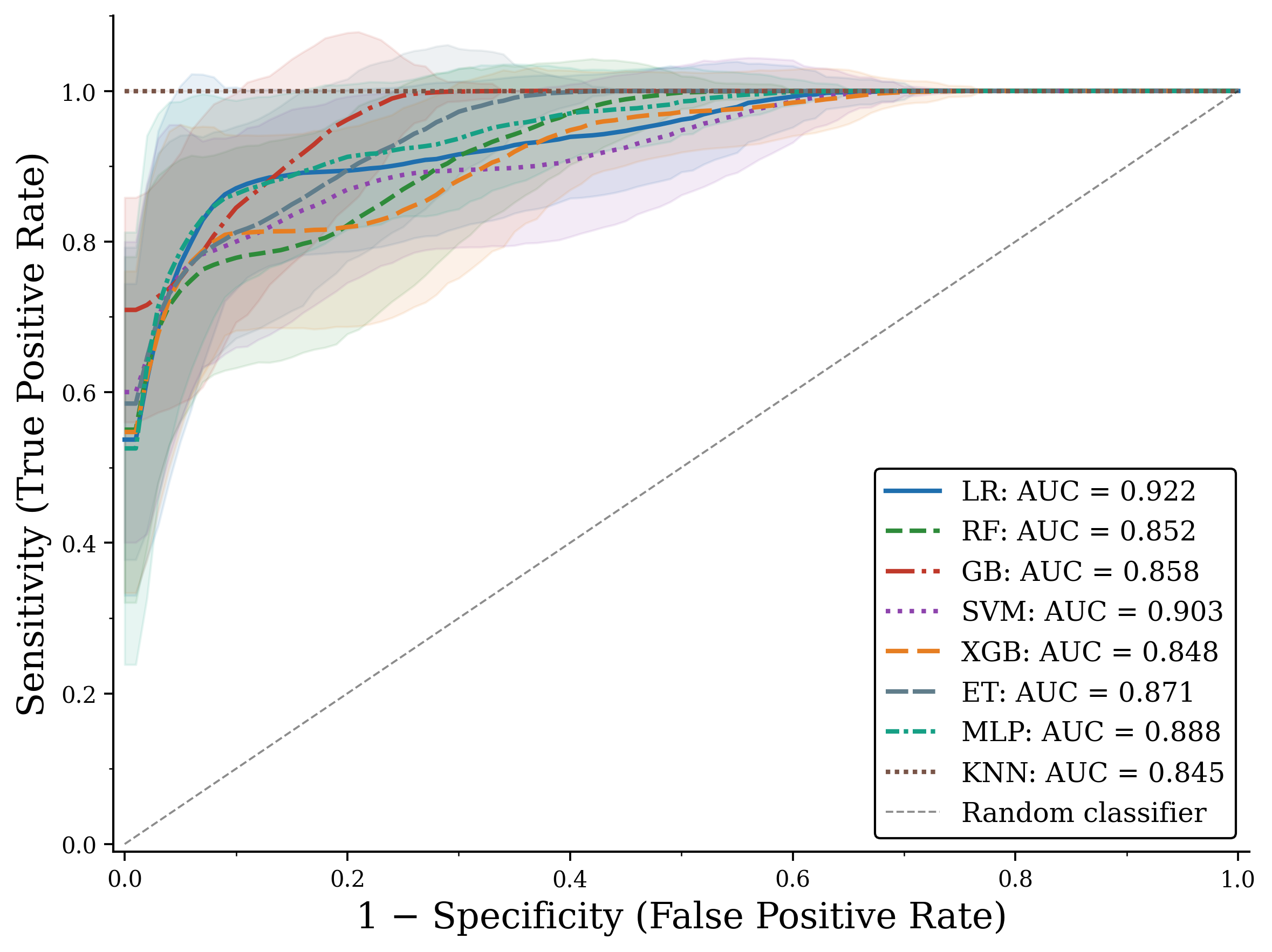}
    \caption{Receiver operating characteristic (ROC) curves comparing the diagnostic performance of the evaluated machine learning models}
    \label{fig:ROC}
\end{figure}

Using the FEA-derived data with parameters given in Table~\ref{tab:fem_constants}, several machine learning models like logistic regression, random forest, gradient boosting, SVM, XGBoost, extra trees, MLP, and KNN were trained and evaluated across 50 folds using five-fold ten-repeat stratified cross-validation. As presented in Table~\ref{tab:model_performance}, logistic regression achieved the highest AUC of 0.92 $\pm$ 0.06 with an overfitting gap of 0.03, indicating strong and stable generalisation. SVM matched this low gap while attaining the highest specificity (0.94), though at the cost of reduced sensitivity (0.64). XGBoost yielded the highest sensitivity (0.81) but showed poor specificity (0.62) and the highest Brier score (0.14), reflecting poorly calibrated probability estimates. KNN and XGB produced near-zero Brier scores, which are an artifact of overconfident hard-boundary predictions rather than genuine calibration, consistent with KNN's large overfitting gap of 0.22. Logistic regression balanced all metrics such as AUC, F1, overfitting gap, and Brier score, most consistently across the 50 evaluation folds. Figure~\ref{fig:ROC} presents the receiver operating characteristic (ROC) curves for the machine learning models, with the logistic regression model achieving the highest AUC of 0.922, followed by SVM and MLP.

\begin{table*}[htbp]
\centering
\caption{Classification Performance Comparison Across Models}
\label{tab:model_performance}
\begin{threeparttable}
    \begin{tabular}{lccccccccc}
        \toprule
        \textbf{Model} & \textbf{AUC} & \textbf{Accuracy} & \textbf{Sensitivity} & \textbf{Specificity} & \textbf{F1} & \textbf{Overfit Gap} & \textbf{Brier} \\
        \midrule
        LR & $\mathbf{0.92 \pm 0.06}$ & $\mathbf{0.86 \pm 0.07}$ & $0.79 \pm 0.13$ & $0.89 \pm 0.09$ & $\mathbf{0.79 \pm 0.10}$ & $\mathbf{0.03}$ & $0.087$ \\
        SVM & $0.90 \pm 0.07$ & $0.84 \pm 0.06$ & $0.64 \pm 0.18$ & $\mathbf{0.94 \pm 0.07}$ & $0.71 \pm 0.13$ & $\mathbf{0.03}$ & $0.098$ \\
        MLP & $0.88 \pm 0.06$ & $0.83 \pm 0.07$ & $0.72 \pm 0.16$ & $0.89 \pm 0.10$ & $0.73 \pm 0.12$ & $0.05$ & $0.08$ \\
        ET & $0.87 \pm 0.08$ & $0.83 \pm 0.06$ & $0.63 \pm 0.18$ & $0.93 \pm 0.07$ & $0.70 \pm 0.13$ & $0.05$ & $0.098$ \\
        RF & $0.85 \pm 0.07$ & $0.80 \pm 0.05$ & $0.67 \pm 0.15$ & $0.87 \pm 0.08$ & $0.68 \pm 0.10$ & $0.05$ & $0.107$ \\
        GB & $0.85 \pm 0.07$ & $0.81 \pm 0.05$ & $0.68 \pm 0.15$ & $0.87 \pm 0.08$ & $0.69 \pm 0.10$ & $0.10$ & $\mathbf{0.00}$ \\
        KNN & $0.84 \pm 0.07$ & $0.77 \pm 0.08$ & $0.67 \pm 0.17$ & $0.81 \pm 0.11$ & $0.65 \pm 0.12$ & $0.22$ & $\mathbf{0.00}$ \\
        XGB & $0.84 \pm 0.07$ & $0.68 \pm 0.08$ & $\mathbf{0.81 \pm 0.14}$ & $0.62 \pm 0.15$ & $0.62 \pm 0.07$ & $0.08$ & $0.14$ \\
        \bottomrule
    \end{tabular}
    \begin{tablenotes}[para,flushleft]
\item  \textbf{Abbreviations:} LR: Logistic Regression, RF: Random Forest, GB: Gradient Boosting, SVM: Support Vector Machine, XGB: Extreme Gradient Boosting, ET: Extra Trees, MLP: Multi-Layer Perceptron, KNN: K-Nearest Neighbour.
\end{tablenotes}
\end{threeparttable}
\end{table*}

Permutation importance analysis identified the most influential FEA-derived features driving the logistic regression classifier. The highest-ranking features had the greatest impact on model performance, highlighting their importance in differentiating between the two groups. As shown in Figure~\ref{fig:AUC permutation importance}, global storage modulus ($G'_{\text{global}}$) and maximum static displacement were identified as the primary diagnostic predictors, contributing to the largest drops in classification performance when permuted. Besides, secondary viscoelastic descriptors such as the loss tangent ($\tan \delta$) and the viscous index ($G''/G^*$) have a mean decrease of $0.00$, indicating no standalone predictive value within this linear framework.

\section{Discussion}\label{sec4}

In this study, our goal was to show that MRI segmentation alone, without vibration hardware, can recover clinically relevant stiffness contrast between healthy and diseased kidneys. The observed stiffness for kidneys, derived from our simulation pipeline, has good agreement with experimentally observed kidney stiffness from existing literature~\cite{rouviere2011magnetic, kirpalani2017magnetic, guven2022evaluation, gandhi2020magnetic, low2015reliability}. The results also showed reduced stiffness in diseased kidneys compared to healthy kidneys, aligning with previous experimental studies~\cite{guven2022evaluation}. 

\begin{figure*}[t]
    \centering
    \begin{minipage}{0.52\linewidth}
        \centering
        \includegraphics[width=\linewidth]{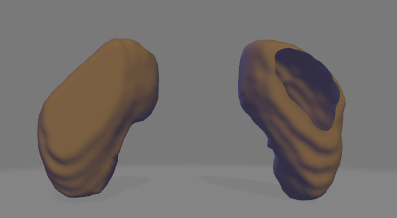}

    \end{minipage}
    \hfill
    \begin{minipage}{0.45\linewidth}
        \centering
        \includegraphics[width=\linewidth]{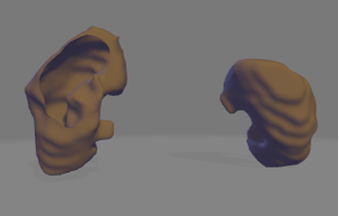}

    \end{minipage}

\caption{Erroneous Surface Topology}
\label{fig:error_surface}
\end{figure*}

Besides, our simulation pipeline ensures that the data driving these top-performing machine learning models is both physically and numerically sound. The high diagnostic accuracy achieved by the optimized logistic regression framework creates a non-invasive computational benchmark for distinguishing healthy controls from chronic kidney disease cohorts. It sets a strong foundation for direct clinical translation and comparative evaluation against existing diagnostic standards.

Despite its robustness, the study has several limitations. Here, we only used patient-specific MRI data. But the structural parameters associated with patients were assumed identical for diseased and healthy kidneys of all patients. Although sensitivity analyses showed that variations in simulation parameters influence the recovered elastographic descriptors, the derived biomechanical features should be interpreted as computationally inferred biomarkers rather than direct estimates of individual tissue properties. Using patient-specific parameters would likely make the parameters more biologically accurate.  

Errors in the initial binary mask directly propagate to the surface mesh. Besides, Gaussian smoothing and Taubin filtering may suppress clinically relevant microstructures. Marching cubes may produce non-manifold edges or disconnected patches in regions of high curvature or noise. This can be seen in errors noted in Figure~\ref{fig:error_surface}. Due to such topological issues, 79 of 100 kidneys in the healthy control group and 50 of 100 kidneys in the chronic kidney disease group were found to be watertight. Another limitation is that for diseased kidneys, both the left and right kidneys were assumed to be affected. However, this labeling is patient-specific, not kidney-specific, which can result in a healthy kidney being marked as diseased.

Besides, the choice of classification angle $\theta$  affects surface partitioning and can lead to inconsistent mesh topology. A total of 48 kidneys from patients with chronic kidney disease and 79 kidneys from healthy controls showed successful volumetric mesh creation with our proposed algorithm. There is also a resolution trade-off. Smaller $h_{\min}$ improves accuracy but increases computational cost $N_e \propto h^{-3}$. A similar effect arises for first-order elements (Tet4), which may not capture complex stress or diffusion gradients accurately compared to higher-order elements. For simplicity, we assumed a linear and isotropic material. However, real kidney tissue exhibits nonlinear and anisotropic behavior not captured here. The Kelvin-Voigt model was simplified, and frequency-dependent viscoelasticity is not fully represented. There are finite-domain effects in which reflections and limited wavelength coverage may bias modulus estimation.

Despite these limitations, the proposed framework demonstrates the feasibility of computational renal elastography from conventional MRI. This method could be a potential alternative to MRE in both research and clinical settings where MRE is not accessible. 
	
\section{Conclusions}\label{sec5}

We developed a computational framework for renal elastographic analysis that combines MRI-derived finite element modeling with viscoelastic inversion to estimate biomechanical properties from conventional kidney MRI. The proposed approach demonstrated that FEM-derived biomechanical parameters capture disease-associated changes and can differentiate healthy kidneys from those with chronic kidney disease with strong classification performance. By integrating physics-based modeling with machine learning, the framework provides interpretable biomechanical biomarkers without requiring dedicated MRE acquisitions.

Although further validation against clinical MRE, histopathology, and patient-specific tissue properties is needed, these findings demonstrate the feasibility of computational renal elastography using routinely acquired MRI. This framework provides a foundation for noninvasive biomechanical assessment of kidney disease and supports future translation through patient-specific modeling, advanced constitutive formulations, and clinical validation.


\printcredits

\section*{Declaration of competing interest}
	The authors affirm that they have no financial conflicts of interest or personal affiliations that could have potentially influenced the research findings or interpretations presented in this study.

\section*{Data availability}
	Data associated with this study are available at: \url{https://github.com/alexdaniel654/Renal_Segmentor}. Code is available upon request from the corresponding author.

\section*{Acknowledgement}
The authors gratefully acknowledge Professors Bachir Taouli and Xiaohong Joe Zhou for their valuable discussions.




\bibliographystyle{elsarticle-num}

\bibliography{cas-refs}





\end{document}